\documentclass[11pt,a4paper]{article}
\pdfoutput=1
\usepackage[utf8]{inputenc}
\usepackage{amsmath,amssymb,amsfonts}
\usepackage{fancyvrb}
\usepackage{xspace}
\usepackage[svgnames]{xcolor}
\usepackage{comment}
\usepackage{enumerate}
\usepackage{multirow}
\usepackage{siunitx}
\usepackage[backend=bibtex]{biblatex}
\usepackage{dmbiblatex}
\usepackage{microtype}
\usepackage{hyperref}

\newcommand{\apron}{\textsc{Apron}}
\newcommand{\coef}{\vec{a}}
\newcommand{\cons}{C}
\newcommand{\coq}{\textsc{Coq}}
\newcommand{\cst}{b}
\newcommand{\defas}{\stackrel{\mbox{\tiny def}}{=}}
\newcommand{\gmp}{\textsc{GMP}}
\newcommand{\impl}{\Rightarrow}
\newcommand{\incl}{\mathop{\sqsubseteq}}
\newcommand{\inclb}{\mathop{\widehat\sqsubseteq}}
\newcommand{\join}{\sqcup}
\newcommand{\joinb}{\widehat\sqcup}
\newcommand{\meet}{\sqcap}
\newcommand{\meetb}{\widehat\sqcap}
\newcommand{\mini}[1]{#1_\text{min}}

\newcommand{\ocaml}{\textsc{Ocaml}}
\newcommand{\pol}{P}
\newcommand{\proj}[2]{\FUN{proj}~#1~#2}

\newcommand{\pt}{\vec{y}}
\newcommand{\satp}[2]{\mbox{Sat}~#1~#2}
\newcommand{\var}{\vec{x}}
\newcommand{\verasco}{\textsc{Verasco}}
\newcommand{\zarith}{\textsc{ZArith}}

\newcommand\MIC[1]{#1}

\newcommand\CST[1]{\textsc{#1}}
\newcommand\FUN[1]{\textit{#1}}

\newcommand\PRG[1]{\texttt{#1}}
\newcommand\RMK[1]{\textcolor{blue}{\textit{\textsf{#1}}}}
\newcommand\UNT[1]{\mathop{\widehat{#1}}}

\def\:={\leftarrow}
\def\*{\times}

\newcommand\checker{\PRG{inclusion\_checker}}
\newcommand\Fv[1]{\Vec{\Lambda}_{#1}}
\newcommand\Fc[1]{\lambda_{#1}}
\newcommand\Fm[1]{F_{#1}}

\newcommand\segp{\FUN{seg}'}
\newcommand\SegpOf[2]{\segp(#1,#2)}
\newcommand\List[1]{[\;#1\;]}

\newcommand\cbar{\alpha}
\newcommand\ACR[1]{\textsc{#1}}
\newcommand\aposteriori{\emph{a posteriori}\xspace}

\newcommand\andl{\,\wedge\,}

\newcommand\COQ{\ACR{Coq}\xspace}

\newcommand\ie{\emph{i.e.}\xspace}

\newcommand\lc{C}

\newcommand\lub{\sqcup}

\newcommand\OCAML{\ACR{Ocaml}\xspace}
\newcommand\Pbar{S_{\!\sss\mathit{bar}}}

\newcommand\projection{\FUN{proj}}
\newcommand\ProjectionOf[2]{\projection~#1~#2}

\let\sss=\scriptscriptstyle

\newcommand\Tuple[1]{(#1)}
\newcommand\vBasic{x'_{i}}
\newcommand\vNonBasic{x'_{n}}
\newcommand\x{\Vec{x}}
\newcommand\xp{\Vec{x}'}

 \DefineVerbatimEnvironment%
   {codeLatex}{Verbatim}
   {commandchars=\\\{\},codes={\catcode`$=3\catcode`_=8\catcode`'=11}}
{\begin{quote}\begin{math}\begin{array}{#1}}
{\end{array}\end{math}\end{quote}}
\newenvironment{math*}{\[}{\]}
\newdimen{\colwidth}
\def\sidebyside{%
\noindent
\colwidth=\textwidth%
\advance\colwidth by -22pt%
\divide\colwidth by2%
\advance\colwidth by 10pt%
\emergencystretch=3cm
\begin{minipage}[t]{\colwidth}\ignorespaces\vspace{0pt}}

\def\nextto{%
 \hfill\end{minipage}\advance\colwidth by -20pt%
  \kern5mm\begin{minipage}[t]{\colwidth}\ignorespaces\vspace{0pt}}%

\def\endsidebyside{\end{minipage}}

{\left\{\begin{array}{@{}c|c@{}}}%
{\end{array}\right\}}

{&\begin{array}{l}}%
{\end{array}}%

\newcommand\coefelt{a}
\newcommand\varelt{x}
\newcommand\compcert{\textsc{CompCert}}
\newcommand\eg{e.g.}
\newcommand\fme{Fourier-Motzkin elimination}
\newcommand\fid{\text{\textit{id}}}
\newcommand\frag{f}
\newcommand{\libpoly}{\textsc{Libpoly}}
\newcommand{\newpolka}{\textsc{NewPolka}}

\newcommand\coefmat{A}
\newcommand\cstvec{\vec b}

\newcommand\fvec{\vec\Lambda}
\newcommand\fvecstr{Farkas vector}
\newcommand\fmat{F}
\newcommand\fmatstr{Farkas matrix}

\newcommand{\coqtyp}[1]{\text{\textbf{#1}}}
\newcommand{\coqval}[1]{\text{$\mathit{#1}$}}

\newcommand{\certtyp}{\mbox{Cert}}
\newcommand{\poltyp}{\mbox{Polyhedra}}

\newcommand{\cert}{cert}

\newcommand{\pararef}[1]{\hyperref[#1]{\S\ref*{#1}}}
\newcommand{\enumref}[1]{\hyperref[#1]{(\ref*{#1})}}

\title{Efficient Generation of Correctness Certificates for the Abstract Domain of Polyhedra\thanks{This work was partially supported by ANR project \href{http://verasco.imag.fr/}{``VERASCO''} (INS~2011).}}

\author{Alexis Fouilhé\thanks{Université Joseph Fourier / VERIMAG; \href{http://www-verimag.imag.fr/}{VERIMAG} is a joint laboratory of \href{http://www.ujf-grenoble.fr/}{Université Joseph Fourier}, \href{http://www.cnrs.fr/}{CNRS} and \href{http://www.grenoble-inp.fr/}{Grenoble-INP}.} \and
\href{http://www-verimag.imag.fr/~monniaux/}{David Monniaux}\thanks{CNRS / VERIMAG} \and
\href{http://www-verimag.imag.fr/~perin/}{Michaël Périn}\footnotemark[1]}

\newcommand{\QQ}{\mathbb{Q}}
\newcommand{\widening}{\triangledown}
\newtheorem{theorem}{Theorem}
\newtheorem{lemma}[theorem]{Lemma}

\newcommand{\narrow}{\fontsize{9pt}{10pt}\selectfont\renewcommand{\tabcolsep}{0.55mm}}
\newcommand{\verynarrow}{\fontsize{10pt}{11pt}\selectfont\renewcommand{\tabcolsep}{0.5mm}}

\bibliography{db}

\begin{document}
\maketitle
\begin{abstract}
Polyhedra form an established abstract domain for
inferring runtime properties of programs using abstract interpretation.
Computations on them need to be certified for the whole static analysis results to be trusted.
In this work, we look at how far we can get down the road of a posteriori verification
to lower the overhead of certification of the abstract domain of polyhedra.
We demonstrate methods for making the cost of inclusion certificate generation negligible.
From a performance point of view,
our single-representation, constraints-based implementation compares
with state-of-the-art implementations.

\end{abstract}

In static analysis by abstract interpretation \cite{cousot77}, sets of reachable states, which are in general infinite or at least very large and not amenable to tractable computation, are over-approximated by elements of an \emph{abstract domain} on which the analyzer applies forward (resp. backward) steps corresponding to program operations (assignments, tests\dots) as well as ``joins'' corresponding to control points with several incoming (resp. outgoing) edges.
When dealing with numerical variables in the analyzed programs, one of the simplest abstract domains consists in keeping one interval per variable, and the forward analysis is known as \emph{interval arithmetic}.
Interval arithmetic however does not keep track of relationships between variables.
The domain of \emph{convex polyhedra} \cite{cousot78} tracks relationships of the form $\sum_i a_i x_i \bowtie b$ where the $a_i$ and $b$ are integer (or rational) constants, the $x_i$ are rational program variables, and $\bowtie$ is $\leq$, $<$ or~$=$.

The implementor of an abstract domain faces two hurdles: the implementation should be reasonably efficient and scalable; it should be reasonably bug-free. As an example, the Parma Polyhedra Library (PPL) \cite{BagnaraHZ08SCP}, version 1.0, which implements several relational numerical abstract domains, comprises 260,000 lines of C++; despite the care put in its development, it is probable that bugs have slipped through. The same applies to the \apron\ library~\cite{jeannet09}.

Such hurdles are especially severe when the analysis is applied to large-scale critical programs (e.g. in the \textsc{Astr\'ee} system \cite{BlanchetCousotEtAl_PLDI03}, targeting avionics software).
For such systems, normal compilers may not be trusted, resulting in expensive post-compilation checking procedures, and prompting the development of the \compcert\ certified compiler~\cite{leroy09}:
this compiler is programmed, specified and proved correct in \coq~\cite{Coq}.
We wish to extend this approach to obtain a trusted static analyzer; this article focuses on obtaining a trusted library for convex polyhedra, similar in features to the polyhedra libraries at the core of PPL and \apron .

One method for certifying the results of a static analysis is to store the invariants obtained by an untrusted analyzer at (roughly) all program points, then check that they are inductive using a trusted checker: each statement is then a Hoare triple that must be checked.
Unfortunately, storing invariants everywhere proved impractical in the \textsc{Astr\'ee} analyzer due to memory consumption; we then opted to recompute them.
Our (future) analyzer will thus store invariants only at loop heads, and thus, for control programs consisting of one huge control loop plus small, unrolled, inner loops, will store only a single invariant.
It will then enter a checking phase which will recompute, in a trusted fashion, all intermediate invariants. Efficiency is thus important.


The main contribution of our article is an efficient way of implementing a provably correct abstract domain of polyhedra. Efficiency is two-fold:
\begin{enumerate}
\item In proof effort: most of the implementation consists in an untrusted oracle providing \emph{certificates} of the correctness of its computations; only a much smaller certificate checker, consisting in simple algorithms (multiplying and adding vectors, replacing a variable by an expression), needs to be proven correct in the proof assistant.
\item In execution time: the expensive parts of the computations (e.g. linear programming) are inside the untrusted oracle and may use efficient programming techniques unavailable in parts that need formal proofs.
We do not compute certificates as an afterthought of polyhedral computations:
close examination of the algorithms implementing the polyhedral operators revealed that they directly expose the elements needed to build certificates.
Simple bookkeeping alleviates the need to rebuild them after the fact.
The overhead of making the operators certifying is thus very limited.
This contrasts with earlier approaches \cite{besson07} based on \emph{a posteriori} generation of witnesses, which had to be recomputed from scratch using linear programming.
\end{enumerate}

A second contribution is a complete implementation of the abstract domain of polyhedra in a purely constraints-based representation.
Most libraries used in static analysis, including PPL and \apron , use a double description: a polyhedron is both an intersection of half-spaces (constraints) or the convex hull of vertices, half-lines and lines (generators), with frequent conversions. Unfortunately, the generator representation is exponential in the number of constraints, including for common cases such as hypercubes (e.g. specification of ranges of inputs for the program).
We instead chose to represent the polyhedra solely as lists of constraints, with pruning of redundant ones. Our implementation uses sparse matrices of rational numbers and uses efficient techniques for convex hull \cite{king05} and emptiness testing by linear programming~\cite{dutertre06}.

We applied our library to examples of polyhedral computations obtained by running the \ACR{Pagai} static analyzer~\cite{henry012} on benchmark programs.
Despite a common claim that implementations based on the double representation are more efficient than those based on constraints only, our library reaches performance comparable to the \ACR{Apron} library together with the high-level of trust brought by our \ACR{Coq} certificate checker.

The remainder of this paper is organized as follows.
After having stated the conventions we are using~(\pararef{defs}),
we define correctness criteria for the operators of the abstract domain~(\pararef{cert}), which all reduce to inclusion properties for which certificates are presented as \emph{Farkas coefficients} (also known as Lagrange multipliers). Such certificates may also be cheaply generated for the convex hull~(\pararef{freecert}). Both forward step and convex hull operations reduce internally to a form of projection.
Some design choices of our implementation are then described~(\pararef{impl}), including how to keep the representation size of the polyhedra reasonable.
Last before conclusion~(\pararef{concl}), an experimental evaluation and accompanying results are presented~(\pararef{exp}).


\section{Definitions and Notations}
\label{defs}
In the remainder of this article, we use the following notations and definitions.
\begin{enumerate}
\item \(\cons\): a linear constraint of the form~$\coef\cdot\var\leq\cst$
	where~\(\coef\) is a vector of rational constants, $\cst$~is a rational
	and~$\var \in \QQ^n$ is the vector of the analyzed program variables.
	Such a linear constraint, or constraint for short, can be viewed as a
	half-space in an \(n\)-dimensional space.
	We write~$\overline{\cons}\defas \coef\cdot\var > \cst$ for the complementary half-space.
\item \(\pol\): a convex polyhedron, not necessarily closed, represented as a set of constraints. 
	We call ``size of the representation of~\(\pol\)'' the number of
	constraints that~\(\pol\) is made of.
\item satisfaction: saying that point~$\pt$ of $\QQ^n$ satisfies a constraint~$\cons\defas \coef\cdot\var\leq\cst$
	means that~$\coef\cdot\pt\leq\cst$.
	By extension, a point~\(\pt\) satisfies~(or is in) polyhedron~\(\pol\) if
	it satisfies all of its constraints.
	We write this:~\(\satp{\pol}{\pt}\).
	Given that a constraint~$\cons$ can be regarded as polyhedron with only one constraint,
	we also write:~$\satp\cons\pt$.
\item Given our focus on the abstract domain of polyhedra we shall adopt the following vocabulary.
	\begin{enumerate}
	\item The order relation on polyhedra~\(\incl\) is geometrical inclusion.
	\item The least upper bound~\(\join\) is the convex hull.
	\item The greatest lower bound~\(\meet\) is geometrical intersection.
	\end{enumerate}
	We will further distinguish the definition of abstract domain operators from
	their actual implementation, which can have bugs.
	The implemented version of the operators will be written with a hat:
	\(\inclb\), \(\joinb\) and~\(\meetb\) implement the ideal operators~\(\incl\), \(\join\) and~\(\meet\), respectively.
\item inclusion: a polyhedron~$\pol_1$ is included in a polyhedron~$\pol_2$
	(noted~$\pol_1\incl\pol_2$) if and only if
	\begin{equation}
	\label{inclprop}
		\forall\pt, \satp{\pol_1}\pt\impl\satp{\pol_2}\pt
	\end{equation}
	Inclusion for constraints~$\cons_1\defas\coef_1\cdot\var\leq\cst_1$
	and~$\cons_2\defas\coef_2\cdot\var\leq\cst_2$ is a special case which is easy to decide:
	$\cons_1\incl\cons_2$ holds if and only if
	there exists $k>0$ such that~$k\cdot\coef_1=\coef_2$ and~$k\cdot\cst_1\leq\cst_2$.
	This latter case is thus proven correct directly inside \coq .
\end{enumerate}


\section{Correctness of the Abstract Domain Operators}
\label{cert}
Let us now see what needs to be proven for the implementation of each operator of an abstract domain
so that the correctness of its result can be established.
\begin{description}
\item[Inclusion test] \(\pol_1\inclb\pol_2\impl\pol_1\incl\pol_2\)
\item[Convex hull] \(\pol_1\incl\pol_1\joinb\pol_2\) and
	\(\pol_2\incl\pol_1\joinb\pol_2\)
	\label{chullcorrect}
\item[Intersection] \(\forall\var,\satp{\pol_1}{\var}\wedge\satp{\pol_2}{\var}\impl\satp{\pol_1\meetb\pol_2}{\var}\).\\
	For now, we will assume a naive implementation of the intersection:
	\(\pol_1\meetb\pol_2\) is the union of the constraints of~\(\pol_1\) with these of~\(\pol_2\), which trivially satisfies the desired property (\ref{inclprop}).
	\label{intersectcorrect}
\item[Assignment] in a forward analysis, $x := e$ amounts to intersection by the equality constraint $x' = e$
(where $x'$ is a fresh variable), projection of $x$ and renaming of $x'$ to $x$.%
\footnote{%
     Other polyhedra libraries distinguish invertible assignments (e.g. $x := x+1$,
more generally $\vec{x'} = A\cdot\vec{x}$ with $A$ an invertible matrix),
which can be handled without projection, from non-invertible ones (e.g. $x := y+z$).
Because our library automatically keeps a canonical system of equalities,
which it uses if possible when projecting, no explicit detection of invertibility is needed;
it is subsumed by the canonicalization.}
When analyzing backward, assignment is just substitution.
\item[Projection] if $\pol_2$ is the returned polyhedron for the projection of $\pol_1$ parallel to variables $x_{i_1},\dots,x_{i_p}$ we check that $\pol_1 \incl \pol_2$ and that variables $x_{i_1},\dots,x_{i_p}$ do not appear in the constraints defining~$\pol_2$.
\item[Widening]: no correctness check needed.
  Widening ($\widening$) is used to accelerate the convergence of the analysis to a candidate invariant.
        For partial correctness of the analyzer, no property is formally needed of the widening operator, since iterations stop when the inclusion test reports that an inductive invariant has been obtained.
        There exist formalizations of the widening operator suitable for proving the total correctness of the analysis (that is, that it eventually converges to an inductive invariant) \cite{Monniaux_HOSC09} but we avoided this question by assuming some large upper bound on the number of iterations after which the analyzer terminates with an error message.
\end{description}

Remark that we only prove that the returned polyhedron \emph{contains} the polyhedron that it should ideally be (which is all that is needed for proving that the results of the analysis are sound), not that it \emph{equals} it: for instance, we prove that the polyhedron returned by the convex hull operator includes the convex hull, not that it is the true convex hull.
The \emph{precision} of our algorithms (that is, the property that they do not return polyhedra larger than needed) is not proved formally; it is however ensured by usual software engineering methods (informally correct algorithms, comparing the output of our implementation to that of other polyhedra libraries\dots).
 
\section{A Posteriori Verification of the Inclusion Test}
\label{incl}
We shall now describe a way to ensure the correctness of the inclusion test.
Recall we represent polyhedra as sets of constraints only.
Our certificate for proving that a polyhedron~\(\pol\), composed of the
constraints~\(\cons_1,\dots,\cons_n\) satisfies a constraint~\(\cons\)
relies on the following trivial fact:
\begin{lemma}
If a point~\(\pt\) satisfies a set of
constraints~\(\left\{\cons_1,\dots,\cons_n\right\}\), it satisfies any linear
positive combination~\(\sum_{i=1}^n\lambda_i\cons_i\) with~\(\lambda_i\geq 0\).
\end{lemma}

If we can find a constraint~$\cons^\prime$  that is a linear positive combination of~\(\cons_1,\allowbreak \dots, \allowbreak\cons_n\) 
and such that~$\cons^\prime\incl\cons$ then
it follows that~\(\pol\) is included in~\(\cons\).
Farkas' lemma states that such linear combinations necessarily exist when inclusion holds, which justifies our approach.

The motivation for \emph{a posteriori} verification of inclusion results stems from
this formulation: while finding an appropriate linear combination requires
advanced algorithms, a small program checking that a particular set of~\(\lambda_i\)'s entails~\(\pol\incl\cons\) can easily be proven correct in a proof assistant.
We call these~\(\lambda_i\)'s the \emph{certificate} for~\(\pol\incl\cons\).
\subsection{A Certificate Checker Certified in \coq}
\label{sec:certified-checker}
Our certificate checker has {\coq} type:
\[
\checker~(\pol_1~\pol_2:\coqtyp{\poltyp})~(\coqval{\cert}:\coqtyp{\certtyp}): \coqtyp{Exception}~(\pol_1\incl\pol_2)
\]
where the type~\(\coqtyp{\poltyp}\) is a simple representation of a polyhedron as a list of linear constraints
and the type~\(\coqtyp{\certtyp}\) is a representation for inclusion certificates.
If a proof of~\(\pol_1\incl\pol_2\) can be built from~\(\coqval{\cert}\),
then the~$\checker$ returns it wrapped in the constructor \CST{value}.
However~\(\coqval{\cert}\) might be incorrect due to a bug in~\(\inclb\).
In this case, the~$\checker$ fails to build a proof of~\(\pol_1\incl\pol_2\)
and returns~\(\CST{error}\).

When extracting the {\ocaml} program from the {\coq} development, proof terms are erased and the type of the checker function becomes that which would have been expected from a hand-written {\ocaml} function:
\footnote{We chose to replace the constructors $\CST{value}$ and $\CST{error}$ of the type \coqtyp{Exception} by {\ocaml} booleans instead of 
letting the extraction define an {\ocaml} type ``exception'' with two nullary constructors due to proof terms being erased.}
\[ 
\checker : \coqtyp{\poltyp} \to \coqtyp{\poltyp} \to \coqtyp{\certtyp} \to \coqtyp{bool}
\]

In reality, our implementation is slightly more complicated because the un\-trust\-ed part of our library, for efficiency reasons, operates on fast rational and integer arithmetic, while the checker uses standard \coq\ types that explicitly represent integers as a list of bits (see~\pararef{sec:coq}).

\subsection{A Certificate-Generating Inclusion Test}
\label{inclalgo}
Let us now go back to the problem of building a proof of~\(\pol\incl\cons\)
by exhibiting an appropriate linear combination.
From~\cite{besson07}, this can be rephrased as a pure satisfiability problem in linear programming:
\[\left(\forall y,\neg\satp{\left(\pol\meet\overline{\cons}\right)}{y}\right) \impl \pol\incl\cons\]
This problem can be solved by the simplex algorithm~\cite{dantzig03}.
For this purpose, the simplex variant proposed by~\cite{dutertre06}, designed for SMT-solvers, is particularly well-suited.
This algorithm only implements the first of the two phases of the simplex
algorithm: finding a feasible point, that is a point satisfying all the
constraints of the problem.
If there is no such point, a witness of unsatisfiability is extracted as a
set of mutually exclusive bounds on linear terms and suitable Farkas coefficient~$\vec{\lambda}$, in the same way that blocking clauses for theory lemmas are obtained for use in SMT-solving modulo linear rational arithmetic.
Furthermore, this algorithm is designed for cheap backtracking (addition and removal of constraints), which is paramount in SMT-solving and also very useful in our application~(\pararef{intersect}).

Our approach to certificate generation differs from previous suggestions \cite{besson07} where inclusion is first tested by untrusted means, and, if the answer is positive, a vector of Farkas coefficients is sought as the solution of a dual linear programming problem with optimization, which has a solution, the Farkas coefficients, if and only if the primal problem has no solution.
Ours uses a primal formulation without optimization.

\subsection{From an Unsatisfiability Witness to an Inclusion Certificate}
\label{unsat}
Inclusion certificates are derived from unsatisfiability witness in a way similar to~\cite{Necula_Lee_CADE2000}.
To illustrate how they are built as part of the inclusion test,
a global idea of the inner workings of the simplex variant from~\cite{dutertre06} is needed.
We insist on the following being a coarse approximation.

We aim at building, given~$\pol$ non-empty and~$\cons$, an inclusion certificate for~$\pol\incl\cons$, otherwise said $\pol\wedge\overline{\cons}$ having no solution.
$\pol$~is composed of~$n$ constraints~$\cons_1,\dots,\cons_n$
of the form~$\sum_{j=1}^n\coefelt_{ij}\cdot\varelt_j\leq\cst_i$, where~$i$ is the constraint subscript.
We refer to~$\overline{\cons}\defas~\cst_0 < \sum_{j=1}^n\coefelt_{0j}\cdot\varelt_j$ as~$\cons_0$.

Let us start by describing the organization of data.
Each constraint~$\cons_i$ is split into
an equation~$\varelt_i'=\sum_{j=1}^n\coefelt_{ij}\cdot\varelt_j$
and a bound~$\varelt_i'\leq\cst_i$ where $\varelt_i'$ is a fresh variable.
For the sake of simplicity, in this presentation,  a constraint $\varelt_i\leq\cst_i$ is represented as $\varelt'_i=\varelt_i \wedge \varelt'_i\leq\cst_i$ ; the actual implementation avoids introducing such extra variable.
Therefore, each~$\varelt_i'$ uniquely identifies~$\cons_i$ by construction
and the original variables $\varelt_i$ are unbounded.
We call \emph{basic} the variables which are defined by an equation (\ie on the left-hand side, with unit coefficient) and \emph{non-basic} the others.
Last, the algorithm maintains a candidate feasible point,
that is a value for every variable~$\varelt_i'$ and~$\varelt_j$, initially set to~$0$.

From this starting point, the algorithm iterates pivoting steps while ensuring 
preservation of the invariant:
\emph{the candidate feasible point always satisfy the equations and the values of the non-basic variables always satisfy their bounds ($\ddagger$).}
At each iteration and prior to pivoting,
a basic variable~$\vBasic$ is chosen such that its value does not satisfy its bounds.
Either there is no such $\vBasic$, and
the candidate feasible point is indeed a solution of~$\pol\wedge\overline{\cons}$,
thereby disproving~$\pol\incl\cons$;
or there is such a basic variable $\vBasic$. In this case, we shift its value to fit its bounds
and
we seek a non-basic variable~$\vNonBasic$ such that its value can be adjusted to compensate the shift: through a pivoting step, $\vBasic$~becomes non-basic, and $\vNonBasic$~becomes basic.
If there is no such $\vNonBasic$
(because all the non-basic variables already have reached their bound),
the equation which defines $\vBasic$ exhibits incompatible bounds of the problem
and is of the form~{$\varelt_i'=\sum_{j\neq i}\lambda_j\cdot\varelt_j'$}
(only~$\varelt_j'$'s appear in this equation: recall that the~$\varelt_j$'s are unbounded).
We now show how to transform this unsatisfiability result into an inclusion certificate.

Since we supposed that~$\pol$ is not empty,
the unsatisfiability necessarily involves $C_0$. Thus,
$\varelt_0'$, which represents~$\cons_0$, has a non-zero coefficient~$\lambda_0$ in the equation.
Without loss of generality,
we suppose that the incompatible bounds involve
an upper bound on~$\varelt_i'$
and that~$\lambda_0$ is positive.
The above equation can be rewritten so that~$\varelt_0'$
appears on the left-hand side:
$$
	\varelt_0'=\sum_{j=1}^n \lambda_j'\cdot\varelt_j'
$$
where the lower bound~$\cst_0<\varelt_0'$ and the upper bound~$\sum_{j=1}^n \lambda_j'\cdot\varelt_j'\leq\cst'$ are such that~$\cst'\leq\cst_0$.
Recall that the~$\varelt_i'$'s were defined as equal to linear terms~$l_i\defas\sum_{j=1}^n\coefelt_{ij}\cdot\varelt_j$ of the constraints $C_i$.
Let us now substitute the~$\varelt_i'$'s by their definition, yielding
$$
	l_0 = \sum_{j=1}^n \lambda_j'\cdot l_j
$$
Noting that~$\cons$ is~$l_0\leq\cst_0$ (since~$\cons_0=~\cst_0<l_0$ is~$\overline{\cons}$),
that~$\sum_{j=1}^n \lambda_j'\cdot l_j\leq\cst$ and
that~$\cst'\leq\cst_0$,
the~$\lambda_j'$'s form an inclusion certificate for~$\pol\incl\cons$.


\section{A Posteriori Verification of the Convex Hull}
\label{freecert}
We saw in~\pararef{chullcorrect} that the result of the convex hull of two polyhedra~\(\pol_1\) and~\(\pol_2\)
must verify inclusion properties with respect to both~\(\pol_1\) and~\(\pol_2\).
Computing~\(\pol\defas\pol_1\joinb\pol_2\), then~\(\pol_1\inclb\pol\) and~\(\pol_2\inclb\pol\)
and then checking the certificates would produce a certified convex hull result, at the expense of two extra inclusion tests.
From a development point of view, this is the lightest approach.
However, careful exploitation of the details of~\(\joinb\) can save us the extra cost of certificate generation,
at the expense of some development effort.

Before delving into the details, let us introduce some more notations for the sake of brevity.
In this section, a polyhedron~$\pol$ is regarded as a column vector of the
constraints~$\cons_1,\dots,\cons_n$ it is composed of.
This allows for a matrix notation:
$\pol\defas \left\{ \var \mid \coefmat\cdot\var\leq\cstvec\right\}$,
where the linear term of~$\cons_i$ is the~$i^\text{th}$ line of~$\coefmat$ and
the constant of~$\cons_i$ is the~$i^\text{th}$ component of~$\cstvec$.

Then, an inclusion certificate, $\lambda_1,\dots,\lambda_n$, for~$\pol\incl\cons^\prime$ is
a line vector~$\fvec$, such that~$\fvec\cdot\pol=\cons$ and~$\cons\incl\cons^\prime$.
Now, an inclusion certificate for~$\pol\incl\pol^\prime$ is a set of inclusion certificates~$\fvec_1,\dots,\fvec_n$,
one for each constraint~$\cons^\prime_i$ of~$\pol^\prime$.
Such a set can be regarded as a matrix~$\fmat$ such that
$$
	\fmat\defas \left(\begin{array}{c} \fvec_1 \\ \vdots \\ \fvec_n \end{array}\right)
	\text{ and } \fmat \* \pol\incl \pol^\prime
$$
where the~$i^\text{th}$ line of~$\fmat\*\pol$ is a constraint~$\cons$
such that~$\cons\incl\cons^\prime_i$.
We call~$\fvec$ a \emph{\fvecstr} and~$\fmat$ a \emph{\fmatstr}.
\subsection{A Convex Hull Algorithm on Constraints Representation}
\label{sec:convexhull}
%
The convex hull~$\pol_1\join\pol_2$ is the smallest polyhedron containing all line segments joining $P_1$ to $P_2$.
Thus, a point $\x$ of $P_1\lub P_2$ is the barycenter of a point $\x_1$ in $P_1$ and a point $\x_2$ in $P_2$.
Exploiting this remark, \cite{benoy05} defined~$\pol_1\join\pol_2$,
with $\pol_i = \{\var \mid \coefmat_i\cdot\var\leq\cstvec_i\}$, as the set of solutions of the constraints
%
%
\begin{math}
A_1 \cdot \Vec{x_1} \leq \Vec{b_1}
\andl
A_2 \cdot \Vec{x_2} \leq \Vec{b_2}
\andl
\Vec{x} = \cbar_1 \cdot \Vec{x_1} + \cbar_2 \cdot \Vec{x_2}
\andl
\cbar_1 + \cbar_2 =1
\andl
0\leq \cbar_1
\andl
0\leq \cbar_2
\end{math}
using $2n+2$ auxiliary variables $\Vec{x_1}, \Vec{x_2}, \cbar_1, \cbar_2$ where $n =|\Vec{x}|$ is the number of variables of the polyhedron.
Still following~\cite{benoy05}, the variable changes $\xp_1 = \cbar_1\cdot\Vec{x_1}$ and $\xp_2 = \cbar_2 \cdot \Vec{x_2}$ remove the non-linearity of the equation $\Vec{x} = \cbar_1 \cdot \Vec{x_1} + \cbar_2\cdot \Vec{x_2} $.

The resulting polyhedron can regarded as the 3-block system~$\Pbar$ below.
The auxiliary variables $\xp_1, \xp_2, \cbar_1, \cbar_2$ are then projected out to stick to the tuple $\Vec{x}$ of program variables.
Therefore, the untrusted convex hull operator $\UNT{\lub}$ mainly consists in a sequence of projections: 
\begin{math}
{P_1} \UNT{\lub} {P_2}  
\defas 
\UNT{\projection} 
~ \Pbar
~\Tuple{\xp_1, \xp_2, \cbar_1, \cbar_2}
\end{math}
where
\[
\Pbar =
\left(\begin{array}{c}
\left.\fbox{$A_1{\xp_1} \leq \cbar_1\Vec{b_1}$} \right.
\\[1ex]
\left.\fbox{$A_2 {\xp_2} \leq \cbar_2 \Vec{b_2}$}\right.
\\[1ex]
\left.\fbox{\scriptsize$
\begin{array}{c}
\x = \xp_1 + \xp_2
\\
\cbar_1 + \cbar_2 =1
\\
0\leq \cbar_1
\\
0\leq \cbar_2
\end{array}$}\right.
\end{array}\right)
\]
\subsection{Instrumenting the Projection Algorithm}
\label{sec-certifying-projection}
Projecting a variable~$\var_k$ from a polyhedron~$\pol$ represented by constraints
can be achieved using \fme~(e.g. \cite{dantzig03}).
This algorithm partitions the constraints of~$\pol$ into three sets:
$E_{x_k}^0$~contains the constraints where the coefficient of~$\var_k$ is nil,
$E_{x_k}^+$~contains those having a strictly positive coefficient for~$\var_k$ and
$E_{x_k}^-$~contains those which coefficient for~$\var_k$ is strictly negative.

Then, the result~\(\pol_{\text{proj}}\) of the projection of~\(\var_k\) from~\(\pol\) is defined as
$$
\pol_{\text{proj}}  = \FUN{proj}~P~\var_k \defas
	E^0_{\var_k}\cup \left(\FUN{map} ~ \FUN{elim}_{\var_k} ~ (E^+_{\var_k}\times E^-_{\var_k}) \right)
$$
where~\(E_{\var_k}^+\times E_{\var_k}^-\) is the set of all possible pairs of inequalities,
one element of each pair belonging to~\(E_{\var_k}^+\) and the other belonging to~\(E_{\var_k}^-\).
The~\(\FUN{elim}_{\var_k}\) function builds the linear combination with positive coefficients of
the members of a pair such that~\(\var_k\) has a zero coefficient in the result.

Illustrating on an example, projecting~$x$ from
$$\pol\defas \{y\leq 1, 2\cdot x + y \leq 2, -x - y \leq 1\} \text{ gives}$$
$$E_x^0 = \{y\leq 1\} \text{ and } E_x^+\times E_x^- = \{ (2\cdot x + y\leq 2, -x -y\leq 1)\}$$
From~$1\cdot (2\cdot x + y \leq 2) + 2\cdot (-x -y\leq 1) = -y\leq 4$,
we get~$\pol_{\text{proj}} = \{y\leq 1, - y\leq 4\}$.

Note that every constraint~$\cons$ of~$\pol_\text{proj}$ is either a constraint of~$\pol$,
or the result of a linear combination with non-negative coefficients~$\lambda_1, \lambda_2$
of two constraints~$\cons_1$ and~$\cons_2$ of~$\pol$, such that
$ \lambda_1\cdot\cons_1 + \lambda_2\cdot\cons_2 = \cons $.
It is therefore possible, with some bookkeeping, to build a matrix~$\fmat$ such that
$\fmat \times \pol = \pol_\text{proj}$.
This extends to the projection of several variables:
if $\proj{\pol}{\var_k} = \pol_\text{proj} = \fmat \times \pol$
and $\proj{\pol_\text{proj}}{\var_l} = \pol_\text{proj}^\prime = \fmat^\prime \times \pol_\text{proj}$,
then $\pol_\text{proj}^\prime = \fmat^{\prime\prime} \times \pol$
with $\fmat^{\prime\prime} = \fmat^\prime \times \fmat$.

\fme\ can generate a lot of redundant constraints,
which make the representation size of~$\pol_\text{proj}$ unwieldy.
In the worst case, the $n$ constraints split evenly into $E_{x_k}^+$ and $E_{x_k}^-$, and thus, after one elimination, one gets $n^2/4$ constraints; this yields an upper bound of $n^{2^p}/4^p$ where $p$ is the number of elimination steps.
Yet, the number of true faces can only grow in single exponential \cite[\S4.1]{Monniaux_CAV10};
thus most generated constraints are likely to be redundant.

The algorithm inspired from~\cite{king05}, which we use in practice, adds these refinements to \fme :
\begin{enumerate}
\item Using equalities when available to make substitutions.
	A substitution is no more than a linear combination of two constraints,
	the coefficients of which can be recorded in~$\fmat$.
	Note that there is no sign restriction on the coefficient applied to an equality.
\item Discarding trivially redundant constraints.
	The corresponding line~$\fmat$ can be discarded just as well.
\item Discarding constraints proved redundant by linear programming, as in \pararef{intersect}.
\end{enumerate}
Note that, since discarding a constraint only \emph{adds} points to the polyhedron, there is no need to prove these refinements to be correct or to provide certificates for them. We could thus very easily add new heuristics.

\subsection{On-the-Fly Generation of Inclusion Certificates}
\label{sec:convex-hull}
In order to establish the correctness of static analysis, the convex hull operator should return a superset of the true convex hull; we thus need proofs of $P_1\incl P_1\UNT\lub P_2$ and $P_2\incl P_1\UNT\lub P_2$. The converse inclusion is not needed for correctness, though we expect that it holds; we will not prove it.
A certifying operator $\UNT{\lub}$ must then produce for each constraint $\lc$ of $P_1\UNT\lub P_2$ 
a certificate $\Fv{1}$ (resp. $\Fv{2}$) proving the inclusion of $P_1$ (resp. $P_2$) into the single-constraint polyhedron $\lc$.
The method we propose for on-the-fly generation of a correctness certificate is based on the following remark.

For each constraint $\lc$ of $P_1\UNT{\lub}P_2$, the projection operator $\UNT{\projection}$ provides a vector $\Fv{}$ such that $\Fv{}\times \Pbar = \lc$,
where $\Pbar$ is the system of constraints defined in \pararef{sec:convexhull}.
An examination of the certificate reveals that $\Fv{}$ can be split into three parts $(\Fv{1},\Fv{2},\Fv{3})$ 
such that 
$\Fv{1}$ refers to the constraints $A_1.{\xp_1} \leq\cbar_1 \Vec{b_1}$ derived from $P_1$ ;
$\Fv{2}$ refers to the constraints $A_2.{\xp_2} \leq\cbar_2 \Vec{b_2}$ derived from $P_2$
and
$\Fv{3}$ refers to the barycenter part $\Vec{x} = {\xp_1} + {\xp_2} \andl \cbar_1+\cbar_2=1 \andl 0\leq \cbar_1 \andl 0\leq \cbar_2$.
Let us apply the substitution $\sigma = [\cbar_1/1, \cbar_2/0,\xp_1/\x, \xp_2/\Vec{0}]$, that characterizes the points of $P_1$ as some extreme barycenters, to each terms of the equality $\Fv{} \* \Pbar = \lc$.
This only changes $\Pbar$: Indeed, $\Fv{}\sigma=\Fv{}$ since $\Fv{}$ is a constant vector and $\lc\sigma=\lc$ since none of the substituted variables appears in $\lc$ (due to projection). 
%
%
We obtain the equality (below) where many constraints of $\Pbar\sigma$ became trivial.
\[
(\Fv{1},\Fv{2},\Fv{3}) 
\times
\left(\begin{array}{c}
\left.\fbox{$A_1{\x} \leq \Vec{b_1}$} \right.
\\[1ex]
\left.\fbox{$0 \leq 0$}\right.
\\[1ex]
\left.\fbox{\scriptsize$
\begin{array}{c}
\x = \x
\\
1 =1
\\
0\leq 1
\\
0\leq 0
\end{array}$}\right.
\end{array}\right)
= 
\lc
\]

This equality can be simplified into $\Fv1 \* (A_1\x\leq\cstvec_1) + \Fc{} (0\leq 1) = \lc$
where $\Fc{}$ is the third coefficient of $\Fv{3}$.
This shows  that $\Fv{1}$ is a certificate\footnote{%
The shift $\Fc{}$ of the bound is lost and will be computed again by our \coq-certified checker.}
for~$\pol_1 \incl\lc$. 
The same reasoning with $\sigma = [\cbar_1/0, \cbar_2/1,\xp_1/\Vec{0}, \xp_2/\x]$ shows that $\Fv{2}$ is a certificate for $\pol_2\incl\lc$.


\section{Notes on the Implementation}
\label{impl}
The practical efficiency of the abstract domain operators is highly sensitive to implementation details. Let us thus describe our main design choices.
\subsection{Extending to Equalities and Strict Inequalities}
Everything we discussed so far deals with non-strict inequalities only.
The inclusion test algorithm however complements such non-strict inequalities,
which yields strict ones.
Adaptation could have been restricted to the simplex algorithm on which the inclusion test relies,
and such an enhancement is described in~\cite{dutertre06}.
We have however elected to add full support for strict inequalities to our implementation.
Once the addition of two constraints has been defined,
almost no further change to the algorithms we discussed previously was needed.

Proper support and use of equalities was more involving.
As~\cite{king05} points out, equalities can be used for projecting variables.
Such substitutions do not increase the number of constraints, contrary to \fme .
We ended up splitting the constraint set into a set of equalities, each serving as the definition of a variable,
and a set of inequalities in which these variables have been substituted by their definitions.
Minimization~(see \pararef{intersect}) was augmented to look for implicit equalities in the set of inequalities.
Last, testing inclusion of~\(\pol\) in~\(\cons\) was split into two phases:
substituting in~\(\cons\) the variables defined by the equalities of~\(\pol\) and
then using the simplex-based method described earlier without putting the equalities of~\(\pol\) in,
which reduces the problem size.

Inclusion certificates were adapted for equalities.
If~$\pol\incl\cons$, with $\cons\defas~\coef\cdot\var = \cst$, cannot be proven
using a linear combination of equalities,
it is split as $\{\coef\cdot\var \leq \cst , \coef\cdot\var \geq \cst\}$
and~$\pol$ is proven to be included in each separately.
\subsection{Minimization}
\label{intersect}
The intersection~\(\pol_1\meetb\pol_2\) is a very simple operation.
As~\pararef{intersectcorrect} described, a naive implementation amounts to list concatenation.
However, some constraints of~\(\pol_1\) may be redundant with constraints
of~\(\pol_2\).
Keeping redundant constraints leads to a quick growth of the representation
sizes and thus of computation costs.
In addition, one condition for the good operations of widening operators on polyhedra is that there should be no implicit equality in the system of inequalities and no redundant constraint~\cite{bagnara05}.

It is therefore necessary to \emph{minimize} the size of the representation of
polyhedra, that is, removing all redundant constraints, and to have a system of equality constraints that exactly defines the affine span of the polyhedron.
We call~\(\mini{\pol}\) the result of the minimization on~\(\pol\).
The correctness of the result is preserved as long as~\(\mini{\pol}\) is
an over-approximation of~\(\pol\), which means~\(\pol\incl\mini{\pol}\).

First, we check whether~\(\pol\) has points in it using the simplex algorithm from~\pararef{unsat}.
If~\(\pol\) is empty, \(\bot\)~is returned as the minimal representation.
The certificate is built from the witness of contradictory bounds returned by the simplex algorithm.
It is a linear combination which result is a trivially contradictory constraint involving only constants
(e.g. \(0 \leq -1\)) and which, in other words, has no solution.

The next step is implicit equality detection.
It builds on~$\coef\cdot\var\leq\cst ~\wedge~ \coef\cdot\var\geq\cst \Rightarrow \coef\cdot\var =\cst$.
For every~$\cons^\leq \defas \coef\cdot\var \leq \cst$ of~$\pol$ (by definition $\pol \incl \cons^\leq$),
we test whether~$\pol \incl \cons^\geq \defas \coef\cdot\var \geq \cst$.
If the inclusion holds, the certificate of the resulting equality is composed of
a linear combination yielding~\(\cons^\geq\)
and a trivial one, $1\cdot \cons^\leq$, yielding~\(\cons^\leq\).
Once this is done, the representation of~\(\pol\) can be split into a system of equalities~\(\pol_e\)
and a system of inequalities~\(\pol_i\) with no implicit equality.
\(\pol_e\) is transformed to be in echelon form using Gaussian elimination, which has two benefits.
First, redundant equations are detected and removed.
Second, each equation can now serve as the definition of one variable.
The so-defined variables are then substituted in~\(\pol_i\), yielding~\(\pol_i^\prime\).
Although our implementation tracks evidence of the correctness of this process,
it should be noted that the uses of equalities decribed above are standard practice.

At this point, if redundancy remains, it is to be found in~\(\pol_i^\prime\) only.
It is detected using inclusion tests: for every~\(\cons\in\pol_i^\prime\),
if~\(\pol_i^\prime\setminus\left\{\cons\right\}\incl\pol_i^\prime\), \(\cons\)~is removed.
Removing a constraint is, at worst, an over-approximation for which no justification needs to be provided.

All that we describe above involve many runs of the simplex algorithm.
The key point which makes this viable in practice is the following: they are all strongly related
and many pivoting steps are shared among the different queries.
We described~(\pararef{unsat}) the data representation used by the simplex variant we use:
it splits each constraint of $\pol$ in linear term and bound by inserting new variables.
These variables can have both an upper and a lower bound.
Let us now illustrate the three steps of minimization on constraint~$\cons\defas \coef\cdot\var \leq \cst$,
split as~$\varelt' = \coef\cdot\var$ and~$\var'\leq\cst$.
The first step, satisfiability, solves this very problem.
Then, implicit equalities detection checks whether~$\varelt'=\coef\cdot\var$ and~$\varelt'<\cst$ is unsatisfiable.
Last, redundancy elimination operates on~$\varelt'=\coef\cdot\var$ and~$\varelt'>\cst$.

For all these problems, we only changed the bound on~$\varelt'$,
without ever touching either the constraint~$\varelt'=\coef\cdot\var$ or the other constraints of $\pol$.
These changes can be done dynamically,
while preserving the simplex invariant ($\ddagger$ of \pararef{unsat}),
by making sure that the affected~$\varelt'$ is a basic variable.
This remark, once generalized to a whole polyhedron,
enables the factorization of the construction of the simplex problem.
Actually, it is only done once for each minimization.
It is also hoped that the feasible point of one problem is close enough to that of the next problem,
so that convergence is quick.

Minimization also plays an important role in the convex hull algorithm.
We mentioned~(\pararef{sec-certifying-projection}) that projection increases the representation size of polyhedra
and described some simple counter-measures from~\cite{king05}.
When projecting a lot of variables, as is done for computing the convex hull of two polyhedra,
each redundant constraint can trigger a lot of extra computation.
Applying a complete minimization after the projection of each variable mitigates this.
More precisely, only the third of the steps described above is used:
projection cannot make a non-empty polyhedron empty
and it cannot reduce the dimension of a polyhedron, no implicit equality can be created.
\subsection{A More Detailed Intuition on Bookkeeping}
We mentioned in~\pararef{inclalgo} and~\pararef{freecert} that simple bookkeeping makes it possible to build inclusion certificates.
We now give a more precise insight on what is involved, on the example of the projection.

The main change is an extension of the notion of constraint,
which is now a pair~$\left(\frag ,\cons\right)$ of a certificate fragment and
a linear constraint as we presented them so far.
A certificate fragment~$\frag$ is a list of pairs~$\left(n_i, \fid_i\right)$,
$n_i$ being a rational coefficient and~$\fid_i$ a natural number uniquely identifying one constraint of~$\pol$.
The meaning of~$\frag$ is the following
$$
	\sum_i n_i\cdot\cons_{\fid_i} = c, \text{ with } \cons_{\fid_i}\in\pol
	\text{ and } \left(n_i, \fid_i\right)\in\frag
$$

The $\FUN{elim}_{\var_k}$ function introduced in~\pararef{sec-certifying-projection} is extended
to take two extended constraints~$(\frag_1,\cons_1)$ and~$(\frag_2,\cons_2)$,
and return an extended constraint~$(\frag,\cons)$.
Recall that the original $\FUN{elim}_{\var_k}$ chooses~$\lambda_1$ and~$\lambda_2$ such that
the coefficient of~$\var_k$ in the resulting~$\cons$ is nil.
The extended version returns
$(\lambda_1\cdot\frag_1 ~@~ \lambda_2\cdot\frag_2,
\lambda_1\cdot\cons_1 + \lambda_2\cdot\cons_2)$,
where @ is the list concatenation operator and~$\lambda_i\cdot\frag_i$ is a notation for:
$$
	\FUN{map} ~ (\FUN{fun} ~ (n, \fid) \rightarrow (\lambda_i\cdot n, \fid)) ~ \frag_i
$$

The certificate fragment keeps track of how a constraint was generated from an initial set of constraints.
For a single projection~$\proj{\pol}{\var_k}$, the fragments are initialized as $[(1, \fid_\cons)]$
for every constraint~$\cons$ before the actual projection starts.
For a series of projection as done for the convex hull, the initialization takes place before the first projection.
\subsection{Polyhedron Representation Invariants}
The data representation our implementation uses for polyhedra satisfies a number of invariants
which relate to minimality.
\begin{enumerate}[(1)]
\item\label{enum:noimpliciteq} There is no implicit equality among the inequalities.
\item\label{enum:noredundantconstraint} There is no redundant constraint, equality or inequality.
\item\label{enum:commonfactor} In a given constraint, factors common to all the coefficients of variables are removed.
\item\label{enum:echelon} Each equality provides a definition for one variable, which is then substituted in the inequalities.
\item Empty polyhedra are explicitly labeled as such.
\end{enumerate}

\enumref{enum:commonfactor}~helps keeping numbers small, hopefully fitting machine representation, resulting in cheaper arithmetic.
\enumref{enum:noimpliciteq}~implies in particular that if an implicit equality is created when adding a constraint~$\cons$ to a polyhedron~$\pol$, then $\cons$ is necessarily involved in that equality. It follows that the search for implicit equalities can be restricted to those involving newly added constraints.
Because of~\enumref{enum:noredundantconstraint}, the same holds for redundancy elimination:
if~$\cons$ is shown to be redundant, $\pol$ remains unaffected by the intersection.
Furthermore, \enumref{enum:echelon} allows for the reduction of the problem dimension when testing for~$\pol_1\incl\pol_2$.
Once the same variables are substituted in~$\pol_1$ and~$\pol_2$,
only the inequalities need to be inserted in the simplex problem.
Last, \enumref{enum:commonfactor}~and~\enumref{enum:echelon} give a canonical form to constraints,
which make syntactic criteria for deciding inclusion of constraints more powerful.
These criteria, suggested by~\cite{king05}, are used whenever possible in the inclusion test and the projection.
\subsection{Data Structures}
\subsubsection{Radix Trees}
Capturing linear relations between program variables with polyhedra generally
leads to sparse systems, as noted by~\cite{king05}.
Our implementation uses a tree representation of vectors\footnote{The idea was
borrowed from~\cite{besson07}.} where the path from the root to a node
identifies the variable whose coefficient is stored at that node.
This offers a middle ground between dense representation, as used by other
widely-used implementation of the abstraction domain of polyhedra, and
sparse representation which makes random access costly as sparsity diminishes.

\subsubsection{Numbers Representation}
Rational vector coefficients can grow so as to overflow native integer representation
during an analysis.
Working around this shortcoming requires the use of an arithmetic library for
arbitrarily large numbers.
This has a serious impact on overall performance.
Our implementation uses the \zarith\cite{zarith}\ \ocaml\ front-end to
\gmp\cite{gmplib}.
\zarith\ tries to lower the cost of using \gmp\ by using native integers as long
as they don't overflow.

Our experiments show that, in many practical cases, extended precision arithmetic is not used.
This echoes similar findings in SMT-solvers such as Z3 or OpenSMT \cite{Caminha_Monniaux_PAAR2012}: in most cases, extended precision is not used, thus the great importance of an arithmetic library that operates on machine words as much as possible, without allocating extended precision numbers.
In the case of polyhedra, however, the situation occasionally degenerates when the convex hull operator generates large coefficients.


The extracted \OCAML code of \PRG{inclusion\_checker} does not use this efficient representation.
Because of the need for correctness of computations,
the checker instead uses the \COQ representation of numbers (lists of bits),
which is inefficient on numerical computations.
Alternatively, assuming trust in {\zarith} and {\gmp}, it is possible to configure the \COQ extractor to base the checker on {\zarith}.

%
\DefineVerbatimEnvironment%
  {code}{Verbatim}
  {commandchars=\\\{\},fontshape=tt,codes={\catcode`$=3\catcode`^=7\catcode`_=8}}
%
\subsection{A Posteriori Certification vs. Full COQ-Certified Development}
\label{sec:coq}
Even though our library is planned to be used in a \COQ-certified analyzer,
we preferred \aposteriori certification over a fully \COQ-certified development.
%
%
Keeping \COQ only for the development of checkers of external computations 
reduces the development cost and reconciles efficiency of the tool and confidence in its implementation through certificates.

First,
it reduces the proof effort: verifying that a guess is the solution to a problem
involves weaker mathematic arguments than proving correctness and termination of the solver.
To illustrate the simplicity of our \COQ development, Figure~\ref{fig-coq-excerpt} shows some excerpts which are self-explanatory.
%
The last function, \PRG{inclusion\_checker}, is representative of the difficulty of the proofs. 
This function is close to its extraction in \OCAML except that it returns either an \CST{error} or a proof of $P_1\incl P_2$ wrapped in the \CST{value} constructor (Line~38).
In the case where $P_1$ is an empty polyhedron (established by $\mathit{eproof}$)
the proof of inclusion in $P_2$ is built from that proof of emptiness.
The missing proof of Line~38 is done in the interactive prover (Lines 43-45) and automatically placed in the function. 
It consists in an induction on the list of constraints of $P_2$ that shows that the empty polyhedron $P_1$ is included in every constraint of $P_2$.

Our external library acts as an oracle: it efficiently performs the operation and returns a certificate 
which serves two purposes: it can be used to check the correctness of the computations 
but it is also a short cut toward the result. 
For instance, the convex hull $P_1\lub P_2$ is easy to obtain from the complete inclusion certificates
$(\Fm{1},\Vec{\Fc{1}})$ related to $P_1$ or $(\Fm{2},\Vec{\Fc{2}})$ related to $P_2$.
Indeed,
$P_1\lub P_2 = \Fm{1} \cdot P_1 + \Vec{\Fc{1}} = \Fm{2}\cdot P_2 + \Vec{\Fc{2}}$ (see \S\ref{sec:convex-hull}).
This way, the expensive computations that involve numerous calls to our simplex algorithm are done by our \OCAML implementation using {\zarith}
and the result is reflected in \COQ  at the cost of just a matrix product using the \COQ-certified representation of numbers.
If we work in such a manner, 
we never actually have to transfer polyhedra from the untrusted to the trusted side.

From a general point of view,
splitting a tool into an untrusted solver and a correctness checker
makes it more amenable to extensions and optimizations.
\emph{A posteriori} certification has a cost each time the correctness of a result needs to be proved
(only during the last phase of the analysis to ascertain the stability of the inferred properties).
However, it allows optimizations whose correctness would be difficult to prove
and usage of untrusted components (\eg\ \gmp).

\begin{figure}
\begin{code}[fontsize=\footnotesize,frame=single,numbers=left] 
\RMK{From module LinearCsrt:} 
Record LinearCstr: Set := mk \{coefs: Vec; cmp\_op: Cmp; bound: Num\}.

Definition Sat ($c$:LinearCstr) ($x$:Vec) : Prop :=
  denote  (Vec.eval (coefs $c$) $x$) (cmp\_op $c$) (bound $c$).

\RMK{From module List:} 
Inductive Forall (A : Type) (pred : A $\to$ Prop) : list A $\to$ Prop :=
  | \CST{Forall\_nil}: Forall pred nil
  | \CST{Forall\_cons}: $\forall$ ($x$:A) (l:list A),  
                   pred $x$ $\to$ Forall pred l $\to$ Forall pred ($x$ :: l)

\RMK{From module Polyhedra:} 
Definition Polyhedra : Set := list (id * LinearCstr).

Definition Sat ($P$:Polyhedra) ($x$:Vec) : Prop :=
  List.Forall (fun $c$ => LinearCstr.Sat (snd $c$) $x$) $P$.
 
Definition Incl ($P$:Polyhedra) ($C$:LinearCstr) : Prop :=
  $\forall$ $x$:Vec, Sat $P$ $x$ $\to$ LinearCstr.Sat $C$ $x$.

Definition (infix $\incl$) ($P_1$ $P_2$ : Polyhedra) : Prop := 
  $\forall$ $x$:Vec, Sat $P_1$ $x$ $\to$ Sat $P_2$ $x$.
 
Definition CertOneConstraint : Set := list (id * Num)

Inductive Cert : Set :=
 | \CST{incl}: list (id * CertOneConstraint) -> Cert
 | \CST{empty}: CertOneConstraint -> Cert.
 
Lemma Empty\_is\_included: $\forall$ ($P$:Polyhedra) ($C$:LinearCstr), 
  (Empty $P$) $\to$ (Incl $P$ $C$).
 
Definition \checker ($P_1$ $P_2$:Polyhedra) ($cert$:Cert) : Exc($P_1{\incl}P_2$).
refine ( match $cert$ with
         | \CST{incl} $icert$ => checkInclusion $P_1$ $P_2$ $icert$
         | \CST{empty} $ecert$ => match (checkEmptyness $P_1$ $ecert$) with
                         | \CST{value} $\mathit{eproof}$ => \CST{value} \_ \RMK{$\leftarrow$ missing proof}
                         | \CST{error} => \CST{error}
                         end
         end
). \RMK{The missing proof is provided by the following proof script:}
induction $P_2$ with IH;
 exact (List.\CST{Forall\_nil} \_ \_) ;
 exact (List.\CST{Forall\_cons} \_ \_ $c$ \_ (Empty\_is\_included $P_1$ (snd $c$) $\mathit{eproof}$) IH).
Defined.
\end{code}
\caption{Excerpts of our \COQ-certified inclusion checker}
\label{fig-coq-excerpt}
\end{figure}

\section{Experimental Results}
\label{exp}
In order to evaluate the viability of our solution, we compared experimentally our library (referred to as {\libpoly}) with mature implementations.

In addition to the efficiency of the polyhedra computation, we wished to measure the cost of the inclusion checker.
Our approach guarantees that,
if our certificate checker terminates successfully on a given verification,
the result of the operation which produced the certificate is correct.
However, this assertion currently only applies to the polyhedra as known to the \coq\ checker:
a translation occurs between the \ocaml\ representation of numbers, \zarith ,
and their representation in the \coq\ language as lists of bits.
This means that the checker has to compute on this inefficient representation, and thus we wished to ascertain whether the cost was tolerable.%
\footnote{%
An alternative would be to map, at checker extraction time, \coq\ numbers to \zarith\ numbers,
at the expense of having both \zarith\ and \gmp\ in the trusted computing base.
One may consider that we already make assumptions about {\zarith} and {\gmp}: we assume they respect memory safety, and thus will not corrupt the data of the {\ocaml} code extracted from {\coq}, or at least that, if they corrupt memory, they will cause a crash in the analyzer (probably in the garbage collector) instead of a silent execution with incorrect data.
This seems a much less bold assumption than considering that they always compute correctly, including in all corner cases.}

The best approach to evaluating \libpoly\ would have been to rely on it for
building a complete static analyzer.
Although this is our long-term goal, a less demanding method was needed for a
more immediate evaluation.
We chose to compare computation results from \libpoly\ to those of widely
used existing implementations of the abstract domain of polyhedra:
the \newpolka\ library and the PPL.
More precisely, we used them through their \apron\ front end~\cite{jeannet09}.
\subsection{The Method}
\label{expmethod}
As~\cite{Monniaux_CAV09} points out, randomly-generated polyhedra do not give a faithful evaluation:
a more realistic approach was needed.
Because of the lack of a static analyzer supporting both \apron\ and \libpoly ,
we carried out the comparison by logging and then
replaying with \libpoly\ the abstract domain operations done
by the existing \textsc{Pagai} analyzer~\cite{henry012} using \apron .

Technically, logging consists in intercepting calls to the \apron\ shared
library (using the wrap functionality of the GNU linker~\texttt{ld}),
analyzing the data structures passed as operands and
generating equivalent \ocaml\ code for \libpoly .
\newpolka\ and PPL results are logged too, for comparison purposes.
At the end of the analysis, the generated \ocaml\ code forms a complete program
which replays all the abstract domain operations executed by the \newpolka\ library or the PPL
on request of the analyzer.

The comparison was done for the following operations: parallel assignment,
convex hull, inclusion test and intersection on the analysis of the following programs:
\begin{enumerate}
\item \texttt{bf}: the Blowfish cryptographic cipher
\item \texttt{bz2}: the bzip2 compression algorithm
\item \texttt{dbz2}: the bzip2 decompress algorithm
\item \texttt{jpg}: an implementation of the jpeg codec
\item \texttt{re}: the regular expression engine of GNU~\texttt{awk}
\item \texttt{foo}: a hand-crafted program leading to polyhedra with many constraints,
	large coefficients and few equalities
\end{enumerate}
\subsection{Precision and Representation Size Comparison}
The result of each operator we evaluated is a well-defined geometrical object.
For every logged call,
the results from \newpolka, PPL and \libpoly\ were checked for equality (double inclusion).
The certificates generated by \libpoly\ were then systematically checked.
Furthermore, polyhedra have a minimal constraints re\-presentation,
up to the variable choices in the substitutions of equalities.
It was systematically checked
whether \libpoly, \newpolka\ and the PPL computed the same number of equalities and inequalities.
In all the cases we tried, the tests of correctness and precision passed.
It is to be noted that the PPL does not systematically minimize representations:
its results often have redundant constraints.%
\footnote{%
This is due to the lazy-by-default implementation of the operators of the PPL.
Since support for the eager version of the operators has been deprecated in and is being removed from the PPL
(see~\cite{ppldoc}, \S\ A Note on the Implementation of the Operators),
we could not configure the library to have the same behavior as \newpolka .}

Besides giving confidence in the results computed by \libpoly ,
ensuring that our results are identical to those of \newpolka\ or the PPL
lead us to believe that the analyzer behavior would not have been very different,
had it used the results from \libpoly .
There is no noticeable difference between the analyses carried out using \newpolka\ and the PPL.
\subsection{Timing Measurements}
Timing measurements were made difficult because of
the importance of the state of polyhedra in the double representation \newpolka\ and the PPL use.
We were concerned that logging and replaying as described above would be unfair towards these libraries, since it would force the systematic recomputation of generator representations that, in a real analyzer, would be kept internally. We thus opted for a different approach.

We measured the timings for \newpolka\ and the PPL directly inside \textsc{Pagai} by wrapping the function calls between calls to a high precision timer.
We made sure that the overhead of the timer system calls was sufficiently small so as to produce meaningful results.
For \libpoly, timing measurements were done during the replay and
exclude the time needed to parse and rebuild the operand polyhedra.

We present two views of the same timing measurements,
carried out on the programs introduced in~\pararef{expmethod}.
Table~\ref{restabprg} gives, for each benchmark program, the total time spent in each operation of the abstract domain.
Such a table does not inform us of the typical distribution of problem sizes and the relationship between problem size and computation time, thus we compiled~Table~\ref{restabsz} which
shows computation times aggregated according to the ``problem size'', defined as the sum of the number of constraints of all the operands of a given operation.

For the assignment and the convex hull,
all the constraints of the two operands are put together after renaming
and many projections follow.
The inclusion test~$\pol_1\inclb\pol_2$, in the worst case,
solves as many linear programming problems as there are constraints in~$\pol_2$,
but each is of size the number of constraints of~$\pol_1+1$.
Last, the intersection operator minimizes the result of the union of the sets of constraints.
Note that the sums in Table~\ref{restabprg} exclude operations on trivial problems of size zero or one.

\begin{table}[t]\setlength{\belowcaptionskip}{0pt}
\caption{Timing comparison between {\newpolka} (N), PPL (P), {\libpoly} (L) and {\libpoly} with certificate checker (C): total time (in milliseconds) spent in each of the operations; trivial problems are excluded.}
\label{restabprg}

\begin{center}\narrow
\begin{tabular}{|l|S[table-format=4.0]S[table-format=5.1]S[table-format=2.1]|S[table-format=4.0]S[table-format=4.1]S[table-format=3.1]S[table-format=3.1]|S[table-format=2.1]S[table-format=2.1]S[table-format=1.1]S[table-format=1.1]|S[table-format=4.0]S[table-format=5.1]S[table-format=2.1]|}
\hline
prog. & \multicolumn{3}{c|}{assignment} & \multicolumn{4}{c|}{convex hull} &
	\multicolumn{4}{c|}{inclusion} & \multicolumn{3}{c|}{intersection}\\
 & N & P & L & N & P & L & C & N & P & L & C & N & P & L\\
\hline
\texttt{bf} & 3.7 & 11.4 & 0.5 & 3.2 & 1.2 & 2.7 & 2.8 & 0.2 & 0.4 & 0.1 & 0.1 & 10.7 & 13.4 & 1.2\\
\texttt{bz2} & 14.6 & 54.1 & 2.9 & 23.5 & 11.5 & 66.8 & 68.7 & 1.6 & 2.8 & 0.7 & 1.2 & 52.3 & 61.1 & 7.9\\
\texttt{dbz2} & 1618 & 4182 & 83.8 & 1393 & 231.9 & 532.8 & 535.3 & 32.3 & 35.6 & 2.1 & 3.6 & 1687 & 1815 & 28.3\\
\texttt{jpg} & 23.7 & 68.3 & 3.8 & 28.2 & 7.5 & 24.0 & 24.9 & 1.2 & 1.8 & 0.5 & 0.8 & 39.7 & 51.0 & 6.0\\
\texttt{re} & 5.7 & 17.2 & 0.7 & 20.2 & 8.4 & 17.9 & 19.2 & 1.1 & 1.3 & 0.5 & 0.7 & 37.3 & 47.2 & 3.3\\ 
\texttt{foo} & 9.2 & 14.8 & 8.5 & 4.2 & 0.6 & 941.8 & 943.7 & 0.2 & 0.2 & 0.9 & 0.9 & 6.7 & 7.1 & 5.5\\
\hline
\end{tabular}
\end{center}
\end{table}

\begin{table}[t]\setlength{\belowcaptionskip}{0pt}\verynarrow
\caption{Timing comparison between {\newpolka} (N), PPL (P) and {\libpoly} (L). Computation times (in milliseconds) are aggregated according to operation and problem size. (n) is the total number of problems of the size range in the benckmarks.}
\label{restabsz}

\newcommand{\totalnr}[1]{\multicolumn{1}{>{\em}r|}{#1}}
\begin{center}
\begin{tabular}{|l l |S[table-format=4.2]|S[table-format=4.2]|S[table-format=4.2]|S[table-format=4.2]|S[table-format=4.2]|S[table-format=4.2]|S[table-format=2.2]|S[table-format=3.2]|}
\hline
\multicolumn{2}{|r|}{problem size} & \multicolumn{1}{c|}{0--1} & \multicolumn{1}{c|}{2--5} & \multicolumn{1}{c|}{6--10} & \multicolumn{1}{c|}{11--15} & \multicolumn{1}{c|}{16--20} & \multicolumn{1}{c|}{21--25} & \multicolumn{1}{c|}{26--30} & \multicolumn{1}{c|}{31+}\\
\hline
\multirow{3}{*}{assignment}
	& N & 33.8 & 601.8 & 385.4 & 20.9 & 78.3 & 537.4 & 59.5 & 13.1 \\
	& P & 47.5 & 1176 & 519.7 & 87.4 & 247.6 & 2111 & 81.7 & 77.9 \\
	& L & 1.1 & 6.6 & 14.3 & 10.7 & 5.2 & 39.2 & 15.2 & 11.6 \\
	& n & \totalnr{539} & \totalnr{667} & \totalnr{381} & \totalnr{58} & \totalnr{64} & \totalnr{480} & \totalnr{30} & \totalnr{16} \\
\hline
\multirow{3}{*}{convex hull}
	& N & 687.9 & 679.7 & 434.1 & 119.5 & 68.8 & 37.9 & 6.4 & 3.5\\
	& P & 167.5 & 141.0 & 68.4 & 22.8 & 16.8 & 9.2 & 1.9 & 0.9 \\
	& L & 7.0 & 57.1 & 133.7 & 131.2 & 1050 & 106.4 & 50.1 & 27.8\\
	& n & \totalnr{3354} & \totalnr{3373} & \totalnr{1092} & \totalnr{354} & \totalnr{135} & \totalnr{65} & \totalnr{14} & \totalnr{7}\\
\hline
\multirow{3}{*}{inclusion}
	& N & 7.2 & 9.7 & 9.7 & 3.3 & 5.8 & 4.0 & 4.0 & 0 \\
	& P & 6.5 & 12.8 & 10.6 & 4.2 & 7.0 & 3.9 & 3.4 & 0 \\
	& L & 0.6 & 1.6 & 1.3 & 0.5 & 1.0 & 0.3 & 0.1  & 0\\
	& n & \totalnr{1482} & \totalnr{1881} & \totalnr{673} & \totalnr{277} & \totalnr{111} & \totalnr{52} & \totalnr{17} & \totalnr{4}\\
\hline
\multirow{3}{*}{intersection}
	& N & 1389 & 1752 & 52.3 & 27.4 & 1.3 & &  & \\
	& P & 1933 & 1740 & 158.6 & 91.4 & 4.8 & & & \\
	& L & 35.0 & 30.9 & 18.4 & 8.8 & 0.6 &  &  & \\
	& n & \totalnr{11458} & \totalnr{4094} & \totalnr{322} & \totalnr{156} & \totalnr{6} & \totalnr{0} & \totalnr{0} & \totalnr{0}\\
\hline
\end{tabular}
\end{center}
\end{table}

The presented results show that \libpoly\ is efficient on small problems.
Yet, the performance gap between \libpoly\ and the other implementations closes on bigger problems.
This is especially true for the convex hull, which is a costly operation in the constraint representation.
At least part of the difference in efficiency on small problems can be explained by the generality \apron\ provides:
it provides a unified interface to several abstract domains at the expense of an extra abstraction layer
which introduces a significant overhead on small problems.

More generally, the use of \zarith\ in \libpoly\ is likely to lower the cost of arithmetic
when compared to \newpolka\ and the PPL, which use \gmp\ directly.
The \texttt{foo} program illustrates this: the analysis creates constraints with big coefficients,
likely to overflow native number representation.
However, precise measurement of the effect of using \zarith\ would be a hard task.

Last, Table~\ref{restabprg} seems to show that problems are most often of rather small size,
but this may well be due to our limited experimentation means.

In spite of the shortcomings of our evaluation method,
these results seem promising for a constraints-only implementation of the abstract domain of polyhedra.
Some progress still needs to be made on the convex hull side (see \pararef{concl}).
It is also interesting to notice the performance differences between the {\newpolka} and the PPL, despite their design similarities; we ignore their cause.

\subsection{Certificate Checking Overhead}
The certificate checking overhead shown in Table~\ref{restabprg} includes
the translation between \ocaml\ and \coq\ representations.
Inside a certified static analyzer, this overhead could be reduced by only transferring the certificates, as opposed to the full polyhedra, and using them to simulate the polyhedra computations, without bothering to check after every call that the polyhedron inside the {\ocaml} library corresponds to the one inside the certified checker.
In addition to translation costs, there is the general inefficiency of computations on {\coq} integers, which are represented as lists of bits; this is considerably more expensive than using native integers, or even arrays of native integers as GMP would do.

However, it should be noted that the checking of inclusion certificates occurs
only during the final step of the certified static analysis which consists in
verifying that the inferred invariant candidates are indeed inductive invariants
for the program.

Last, the overhead of certificate checking is relatively greater for inclusion than for convex hull.
Although the actual checking burden is bigger for the convex hull,
due to certificate composition densifying the resulting certificate,
the inclusion test algorithm is much cheaper than the convex hull in terms of computations.
More precisely, the convex hull algorithm involves inclusion tests as part of representation minimization.


\section{Conclusions}
\label{concl}
The previous sections demonstrated that a realistic implementation of the abstract domain of polyhedra
can be certified using a posteriori verification of results.
This approach has a key benefit: the time-consuming development inside the \coq\ proof assistant is reduced to the bare minimum.
A tight integration of the certification concern enables on-the-fly certification generation
as a by-production of the actual computations,
thereby making the associated cost negligible.
The same procedures can be used for fixed point iterations (with certificate generation turned off for efficiency) and for fixed point verification (with certificates generated and checked).

The complete implementation which has been developed operates only on a constraints representation of polyhedra; our motivations for this choice were the ease of generation of certificates as well as the absence of combinatorial explosion on common cases such as hypercubes.
This is made possible through careful choice of data structures and
exploitation of recent algorithmic refinements~\cite{king05,dutertre06}.
Possible future developments include designing efficient techniques for generating Farkas certificates for a library based on the double representation (generators and constraints)
and providing heuristics for choosing when to operate over constraints only and when to use the double representation.

Prior to this, however, there remains room for both enhancement and extension of our current implementation.
A simple enhancement would be to have both an upper and a lower bound for linear terms, which would further condense the re\-presentation of polyhedra.
The implicit equality detection algorithm could be made less naive by exploiting the fact that
a point in a polyhedron~$\pol$ which has implicit equalities~$E_i$ necessarily reaches the bounds of the inequalities involved in the proof of~$\pol \incl E_i$.

\MIC{Finally, our library is planned to be part of a certified static analyzer, such as the one being built in the {\verasco} project. Beyond a certified  implementation of the abstract domain of polyhedra, our library could also serve to 
verify the numerical invariants discovered by untrusted analysis using a combination of abstract domains (intervals, octagons, ... 
which are special cases of polyhedra). The discovered invariants could be stored in the form of polyhedra and the verification of their stability
could be done with our certified library.}
Currently, our polyhedron library only deals with linear constraints, but a general-purpose analyzer
has to handle nonlinearity.
Our library should therefore include linearization techniques~\cite{mine06} 
at the condition that these be proven correct. 

\paragraph{Acknowledgements}
We would like to thank Bertrand~Jeannet for his advice on proper ways to evaluate \libpoly\ against his
\newpolka\ library.


\printbibliography[heading=bibintoc]
\end{document}